\documentclass[aps,prd,twocolumn,nofootinbib,superscriptaddress,10pt,floatfix]{revtex4-1}
\usepackage{amstext}
\usepackage{amssymb}
\usepackage{amsmath}
\usepackage{tabularx}
\usepackage{caption}
\usepackage{float}
\usepackage[utf8]{inputenc}
\usepackage{color}
\usepackage{graphicx}
\usepackage{array}
\usepackage{subfigure}
\usepackage[export]{adjustbox}
\usepackage{color}
\usepackage{listings}
\usepackage{natbib}
\usepackage{appendix}

\setcitestyle{unsrtnat,open={[},close={]}}
\usepackage{url}
\usepackage{bm}
\usepackage{multirow}
\usepackage{etoolbox}
\usepackage[utf8]{inputenc}
\usepackage[nolist]{acronym}
\usepackage[breaklinks, plainpages=false, colorlinks=true, anchorcolor=cyan, linkcolor=magenta, citecolor=cyan, urlcolor=purple, bookmarks=false]{hyperref}
\usepackage[normalem]{ulem}
\usepackage{tabu}
\usepackage{tikz}
\usetikzlibrary{arrows,shapes,trees,decorations.pathreplacing}
\usepackage{import}
\usepackage{svn-multi}

\begin{document}

\preprint{APS/123-QED}

\title{Unmasking noise transients masquerading as intermediate-mass black hole binaries}
\author{Sayantan Ghosh}
    \email{stanghosh@iitb.ac.in}
    \affiliation{Department of Physics, Indian Institute of Technology Bombay, Powai, Mumbai 400 076, India}

\author{Koustav Chandra}
    \email{kbc5795@psu.edu}
    \affiliation{Department of Physics, Indian Institute of Technology Bombay, Powai, Mumbai 400 076, India}
    
\author{Archana Pai}
    \email{archanap@iitb.ac.in}
    \affiliation{Department of Physics, Indian Institute of Technology Bombay, Powai, Mumbai 400 076, India}

\date{\today}
      

\begin{abstract}

In the first three observation runs, ground-based \ac{GW} detectors have observed close to $\sim 100$ \ac{CBC} events. The \ac{GW} detection rates for \acp{CBC} are expected to increase with improvements in the sensitivity of the \ac{IGWN}. However, with improved sensitivity, non-Gaussian instrumental transients or ``glitches'' are expected to adversely affect \ac{GW} searches and characterisation algorithms. The most detrimental effect is due to short-duration glitches, which mimic the morphology of short-duration \ac{GW} transients, in particular \ac{IMBH} binaries. They can be easily misidentified as astrophysical signals by current searches, and if included in astrophysical analyses, glitches mislabelled as \ac{IMBH} binaries can affect \ac{IMBH} population studies. In this work, we introduce a new similarity metric that quantifies the consistency of astrophysical parameters across the detector network and helps to distinguish between \ac{IMBH} binaries and short-duration, loud glitches which mimic \ac{IMBH} binaries. We develop this method using a simulated set of \ac{IMBH} binary signals and a collection of noise transients identified during the third observing run of the Advanced LIGO and Advanced Virgo detectors.
\end{abstract}

\maketitle

\acrodef{ROC}[ROC]{Receiver Operating Characteristic}
\acrodef{LR}[LR]{likelihood ratio}
\acrodef{GWOSC}[GWOSC]{Gravitational Wave Open Science Center}
\acrodef{KDE}[KDE]{kernel density estimate}
\acrodefplural{KDE}[KDEs]{kernel density estimates}
\acrodef{BBH}[BBH]{binary black hole}
\acrodef{BH}[BH]{black hole}
\acrodef{LVC}[LVC]{LIGO Scientific and Virgo Collaborations}
\acrodef{GW}[GW]{gravitational wave}
\acrodefplural{GW}[GWs]{gravitational waves}
\acrodef{CBC}[CBC]{compact binary coalescence}
\acrodefplural{CBC}[CBCs]{compact binary coalescences}
\acrodef{CI}[CI]{confidence interval}
\acrodef{IMBH}[IMBH]{intermediate-mass black hole}
\acrodef{SMBH}[SMBH]{supermassive black hole}
\acrodefplural{SMBH}[SMBHs]{supermassive black holes}
\acrodefplural{IMBH}[IMBHs]{intermediate-mass black holes}
\acrodef{SNR}[SNR]{signal-to-noise ratio}
\acrodef{FAR}[FAR]{false alarm rate}
\acrodef{PSD}[PSD]{power spectral density}
\acrodefplural{PSD}[PSDs]{power spectral densities}
\acrodef{LVK}[LVK]{LIGO Scientific, Virgo and KAGRA}
\acrodef{GR}[GR]{General Relativity}
\acrodef{FF}[FF]{fitting factor}
\acrodef{O3}[O3]{third observing run} \acrodef{GWTC}[GWTC]{Gravitational Wave Transient Catalogue}  
\acrodefplural{GWTC}[GWTCs]{Gravitational Wave Transient Catalogues}
\acrodef{IFAR}[IFAR]{inverse false alarm rate}
\acrodef{BHB}[BHB]{black hole binary}
\acrodefplural{BHB}[BHBs]{black hole binaries}
\acrodef{LHO}[LHO]{Hanford}
\acrodef{LLO}[LLO]{Livingston}
\acrodef{JSD}[JSD]{Jensen Shannon divergence}
\acrodef{IGWN}[IGWN]{International Gravitational-Wave Observatory Network}
\acrodef{cWB}[cWB]{coherent waveburst}
\acrodef{PE}[PE]{parameter estimation}
\section{\label{sec:intro} Introduction}

In recent years, interest has surged in the detection and characterisation of \acfp{IMBH}, which are black holes having masses between $10^2 \mathrm{M_\odot}$ and $10^5 \mathrm{M_\odot}$ \cite{imbh-review}. They are believed to be seeds of \acfp{SMBH} ($M > 10^5 \mathrm{M_\odot}$). In general, the observation and characterisation of \acp{IMBH} are expected to lead to a better understanding of how \acp{SMBH} form, which will eventually enable us to study the formation and growth of galaxies. \cite{imbh-review}

The ground-based \ac{GW} detectors Advanced LIGO \cite{adv-ligo-paper} and Advanced Virgo \cite{adv-virgo-paper} are sensitive to \ac{IMBH} binary mergers producing remnants in the lower ($10^2-10^3\mathrm{M_\odot}$) IMBH mass range \cite{o1o2-imbh}, thus opening up new prospects for detecting and characterising \ac{IMBH} binaries. These detectors have completed three observing runs - O1 , O2 and O3 \cite{1-ogc,2-ogc,3-ogc,4-ogc,gwtc-1,gwtc-2,gwtc-3} and the fourth observation run O4 has begun recently. To date, $\sim$ 100 \acf{CBC} events have been detected, most of which are \ac{BBH} mergers. The hallmark of O3 was the first confident \ac{IMBH} binary event - GW190521 \cite{gw190521-discovery,gw190521-properties}. This signal, under the quasi-spherical black hole binary hypothesis, is consistent with the merger between two very massive black holes (masses $85 M_\odot$ and $66 M_\odot$), leaving behind an \ac{IMBH} remnant of mass $\sim 142 M_\odot$. There was a dedicated \ac{IMBH} binary search for the O3 run \cite{o3-imbh, pycbc-imbh, cwb}, which declared GW190521 as the only significant \ac{IMBH} binary event along with a couple of marginal candidates. 

\ac{IMBH} binary signals detectable in the LIGO-Virgo bandwidth have short duration ($<0.1s$), and are dominated by the merger and ringdown phases, with a small contribution to the \acf{SNR} coming from the inspiral part. This makes detecting and characterising \ac{IMBH} binaries challenging. Non-Gaussian, non-stationary noise transients or ``glitches" appearing in the strain data adversely affect search sensitivity and astrophysical studies \cite{o3-imbh}. Many glitch classes are detrimental to \ac{IMBH} searches. Due to their durations ($< 0.1s$) and frequency bandwidths (10 to a few hundred Hz) being similar to those of IMBH binary signals \cite{derek-davis-thesis}, some glitch classes, such as Tomtes and Blips, are prone to be flagged as \ac{IMBH} binary signals by the current detection pipelines. 200214\_224526, one of the marginal candidates detected in the O3 IMBH search, is a classic case of glitches masquerading as an \ac{IMBH} binary event. Although this trigger was detected by the \ac{cWB} pipeline, subsequent investigations revealed that this trigger was of instrumental origin \cite{o3-imbh}. 

It may not always be possible to ascertain whether a particular transient is of instrumental or astrophysical origin. For example, the candidate 200114\_020818 was detected in \ac{LHO}, \ac{LLO} and Virgo detectors with a network \ac{SNR} of 14.5 and \acf{FAR} of $0.06 \mathrm{yr}^{-1}$ by the O3 IMBH search \cite{o3-imbh}. When this candidate was analysed with the three waveform models --- IMRPhenomXPHM \cite{imrphenomxphm-ref}, SEOBNRv4PHM \cite{seobnrv4phm-ref} and NRSur7dq4 \cite{nrsur7dq4-ref}, the posteriors obtained were inconsistent with one another. It is quite likely that the inconsistency between the posteriors is due to waveform systematics. Still, there is also another possibility that this candidate is a culmination of chance coincident glitches in multiple detectors. So, this candidate could neither be confidently ruled out as a glitch nor could it be flagged as a confident \ac{IMBH} binary event. Also, the coherence test did not provide sufficient evidence in support of the coherent signal hypothesis \cite{o3-imbh}. Thus, it is crucial to develop additional independent tests that can increase confidence in deciding the nature of transients of doubtful origin. From an astrophysical standpoint, distinguishing between \ac{IMBH} binaries and \ac{IMBH} binary mimickers is paramount. This is because the incorrectly classified noisy transients can severely affect the astrophysical population studies of IMBH binaries, which are rare events compared to stellar \ac{BBH} mergers.

In order to minimise the effect of glitches, \ac{GW} search algorithms use a variety of distinct classes of vetoes \cite{vetoes}, gating \cite{pycbc-paper} and test statistics that can discriminate between signals and glitches \cite{chisquared}. In the \ac{PE} study, a test was developed in \cite{veitch} that assesses if a given trigger is of noise origin in the multi-detector Bayesian framework. This coherence test computes the Bayes Factor between the hypothesis that the data contains a coherent \ac{CBC} signal across the detectors ($H_{\mathrm{coh}}$) against the hypothesis that the data contains incoherent instrumental features ($H_{\mathrm{inc}}$). The evidence for $H_{\mathrm{coh}}$ is computed by assuming that the incoming astrophysical signal is characterised by the same signal parameter vector $\vec{\theta}$ in all detectors.  The evidence for $H_{\mathrm{inc}}$ is the product of the evidence in the individual detectors, allowing for different, and thus inconsistent parameters across the detector network.

In addition, there have been several attempts to address the problem of distinguishing signals from glitches in the detector characterisation stage. BayesWave approaches this problem in a morphology-independent way, by projecting both signals and glitches onto the wavelet basis \cite{bayeswave,bayeswave-glitch-subtraction}. Parameterised models of frequently occurring glitch types have been developed using probabilistic principal component analysis \cite{glitschen}. Recently, efforts have been made to develop population models of glitches by projecting the glitches into the parameter space for astrophysical models \cite{population-models}.  

In this work, we develop a new, simplistic approach that helps distinguish between \ac{IMBH} binary signals and short-duration glitches. This method assumes that regardless of the waveform model, the posteriors of any astrophysical transient should be
consistent between the detectors. Glitches, being of local origin, may produce inconsistent posteriors. We use \ac{JSD} \cite{jsd-ref} as a similarity metric to quantify the consistency between parameter estimates. We develop this method using a simulation campaign in the \ac{IMBH} binary parameter space and glitches in the PyCBC-IMBH search \cite{pycbc-imbh}. Finally, we apply this method to several high-mass ($M_T^\mathrm{src} > 65\mathrm{M_\odot}$) GW events and marginal candidates in the GW catalogues.

The paper is structured as follows: Section \ref{sec:pe} reviews the basics of Bayesian parameter inference of GW signals. Section \ref{sec:pe_coh} introduces a JSD-based similarity metric that quantifies the parameter consistency between two detectors. Section \ref{sec:sim} describes the astrophysical simulations and glitches analysed in this work. Section \ref{sec:test-statistic} defines the new test statistic used in this work and also summarises the main results. Finally, section \ref{sec:application} discusses the application of the method developed in this work on a curated set of \ac{GW} triggers and section \ref{sec:conclusion} summarises the main conclusions of this work.
    
\section{\label{sec:methods} Methodology}

\subsection{\label{sec:pe} Parameter Estimation in the Bayesian Framework}
The calibrated strain data $d(t)$ in the interferometric \ac{GW} detectors is modelled as:
\begin{equation}
    d(t) = n(t) + s(t;\Vec{\theta}) \, ,
\end{equation}
where, $n(t)$ is the detector noise and $s(t;\Vec{\theta})$ is the GW signal. 
The noise $n(t)$ can be modelled as:
\begin{equation}
    n(t) = n_G(t) + n_{NG}(t) \, , 
\end{equation}
where $n_G(t)$ is Gaussian and wide-sense stationary. $n_{NG}(t)$ denotes glitches, which are non-Gaussian and non-stationary. The posterior of the parameters $\Vec{\theta}$ describing the signal is given by \cite{veitch,thrane}: 
\begin{equation}
    p(\Vec{\theta}|d,W) = \frac{\mathcal{L}(d|W,\Vec{\theta}) \pi (\Vec{\theta}|W)}{\mathcal{Z}} \, , 
\end{equation}
where $d$ is the strain data, $W$ is the waveform approximant which models the signal, $\mathcal{L}(d|W,\Vec{\theta})$ is the likelihood function, which gives the probability of obtaining the data $d$ given the waveform model $W$ and the parameters $\Vec{\theta}$. Under the assumption that the noise is Gaussian, the likelihood takes a Gaussian form. $\pi (\Vec{\theta}|W)$ is the prior distribution on the parameters $\Vec{\theta}$. $\mathcal{Z}$ is called the \textit{model evidence} because it gives the total probability of the model $W$ producing the data $d$, marginalised over all parameters. 
\begin{equation}
    \mathcal{Z} = \int \mathcal{L}(d|W,\Vec{\theta}) \pi (\Vec{\theta}|W) d\Vec{\theta} . 
\end{equation}
The posterior distribution of a single parameter is then obtained by marginalising the multi-dimensional posterior over all the other parameters, also known as \textit{nuisance parameters}.

\subsection{\label{sec:pe_coh}Jensen Shannon Divergence: A measure of two-detector parameter consistency}
An astrophysical signal $s(t;\Vec{\theta})$ (in our case, an \ac{IMBH} binary signal) arrives at two detectors with a time difference consistent with the light travel time between the two detectors. The signal amplitude in each detector is a linear combination of the plus and cross polarisations, weighted by the antenna pattern functions, which give the directional sensitivity of a detector to a particular polarisation. The phase evolution of the signal in each detector depends on the intrinsic parameters, i.e, the masses and spins of the merging black holes. We expect the signal parameters estimated separately from the data in two detectors to be consistent with each other, barring the noise \acf{PSD} effects pertaining to the noise in each detector. This feature should not depend on the waveform model used. Of course, different waveform models might introduce biases, but these biases are also expected to be consistent across the detectors. On the other hand, if there are two non-astrophysical but temporally chance-coincident noise transients($n_{NG}(t)$) in the two detectors which happened to trigger the same template, then the parameters estimated from the strain data in the two detectors (assuming that they are of astrophysical origin) may not be consistent with each other, as local disturbances cause the noisy transients. A matched filter's output is simply the glitch's projection on a particular template and does not need to describe the entire glitch morphology. However, the \ac{PE} samples provide a more complete description of the morphology of the transient. Thus, we expect that the posterior distributions of the astrophysical parameters corresponding to two detectors will be consistent for an incoming astrophysical signal and will be either non-overlapping or minimally overlapping for chance coincident glitches. 

In our approach, we evaluate the likelihoods by assuming that a quasi-circular quadrupole waveform model well describes the transient. Subsequently, we assess the compatibility of the inferred posteriors of astrophysical parameters across the detector network. Unlike the coherence test, which compares the evidence for the coherent and incoherent hypotheses, we use \acf{JSD} \cite{jsd-ref} to quantify how distinguishable the posteriors in different detectors are. This provides us with a \textit{bounded}, \textit{symmetric} measure of the difference between the information content in two probability distributions. It can be computed for both one-dimensional and multi-dimensional distributions. Mathematically, it is defined as follows. Let $\theta$ be one of the parameters describing an \ac{IMBH} binary signal. Let $p_H(\theta)$ and $p_L(\theta)$ be the posteriors of $\theta$ obtained from PE runs with LHO and LLO data respectively.

The \ac{JSD} between $p_H(\theta)$ and $p_L(\theta)$ is defined as \cite{jsd-ref}:
\begin{equation} \label{jsd}
\begin{split}
    \mathrm{JSD}_{\theta}(p_H||p_L) = \frac{1}{2}\int p_H(\theta)\log_2\frac{p_H(\theta)}{p_\mathrm{avg}(\theta)} d\theta + \\
    \frac{1}{2}\int p_L(\theta)\log_2\frac{p_L(\theta)}{p_\mathrm{avg}(\theta)} d\theta  \, ,
\end{split}
\end{equation}
where, 
\begin{equation}
    p_\mathrm{avg}(\theta) = \frac{1}{2}(p_H(\theta) + p_L(\theta))
\end{equation}
We can further extend this definition to two dimensions as follows: 
\begin{equation}{\label{jsd-2d-defn}}
\begin{split}
    \mathrm{JSD}_{\theta_1,\theta_2}^\mathrm{2D}(p_H||p_L) =
    \frac{1}{2}\int p_H(\theta_1,\theta_2)\log_2\frac{p_H(\theta_1,\theta_2)}{p_\mathrm{avg}(\theta_1,\theta_2)} d\theta_1 d\theta_2 \\ + 
    \frac{1}{2}\int p_L(\theta_1,\theta_2)\log_2\frac{p_L(\theta_1,\theta_2)}{p_\mathrm{avg}(\theta_1,\theta_2)} d\theta_1 d\theta_2  \, , 
\end{split}   
\end{equation}
where $\theta_1$ and $\theta_2$ are two parameters. 

For two identical distributions, the \ac{JSD} is zero. The greater the value of the \ac{JSD}, the greater the dissimilarity between the two distributions, with the maximum possible value of unity. For signals, we expect the posteriors corresponding to the two detectors to be very similar to each other, and hence the \ac{JSD} is expected to be close to zero. For chance-coincident glitches, we expect the posteriors corresponding to the two detectors to be dissimilar; hence, the \ac{JSD} is expected to be close to 1. 

\subsection{\label{sec:sim}Simulations}

We simulate 1000 IMBH binary signals with the model NRSur7dq4 \cite{nrsur7dq4-ref}, a quasi-spherical, multipolar waveform model calibrated to numerical relativity simulations. The parameters of these simulated signals are chosen randomly from the following ranges:$ M_T^\mathrm{det} \sim \mathcal{U}(150,500) M_\odot$, mass ratio $\sim \mathcal{U} (1,6)$, the spin magnitudes $\chi_1, \chi_2 \sim \mathcal{U}(0,0.9)$, and the directions of spins are chosen from an isotropic distribution. These signals are added to different realisations of Gaussian noise coloured by the \ac{PSD} of the two Advanced LIGO detectors in the O3 run, with the LHO-LLO network \ac{SNR} $\sim \mathcal{U}(10,20)$. We chose this \ac{SNR} window because most of the observed GW events lie in this range.  

Often, obtaining glitches for a playground study is a challenging task. In GW astronomy, the noise background study is carried out by artificially sliding the data of one detector with respect to the other detector by unphysical time delays. The search algorithm is processed on this data, and the chance-coincident noisy events provide the statistic for the noise background. As a playground, here we consider $\sim 500$ pairs of LHO-LLO glitches in the background data of the O3 PyCBC-IMBH search - an optimised matched-filter-based search for \ac{IMBH} binaries \cite{pycbc-imbh}. The pairs of glitches that we consider were template-coincident as well as temporally coincident in the time-shifted background constructed for estimating the noise statistic for the search. 

We perform \acf{PE} runs with the \ac{LHO} and \ac{LLO} data separately for both signals and glitches with IMRPhenomXAS \cite{imrphenomxas-ref}. We use this waveform model for our recovery as it is a frequency-domain quasi-circular quadrupole waveform model. This allowed us to use likelihoods that are explicitly marginalised over the effective phase, distance and time, thereby reducing the number of dimensions and speeding up likelihood evaluations \cite{thrane}. We understand that for signals, the difference between the injection and recovery models can introduce a bias in the estimates of the parameters. However, we expect this bias to be similar in the two detectors, and it should not affect the \ac{JSD} study between the posteriors. We note that here our goal is not to recover the injected parameters but to assess the consistency between the parameter estimates in the two detectors. By no means do we propose to use these parameter estimates to infer the properties of the underlying signal. 

All \ac{PE} runs are performed with \texttt{PyCBC Inference}~\cite{pycbc-inference-paper}. We use the \texttt{dynesty} sampler \cite{dynesty-paper} and employ an aligned-spin prior, which is uniform in the detector-frame component masses, uniform in the aligned-spin magnitudes, isotropic in sky location and uniform in Euclidean volume. We take a lower frequency cutoff of 15 Hz for likelihood evaluations. 

\begin{figure*}
    \begin{minipage}{\textwidth}
      {\includegraphics[width=0.45\linewidth]{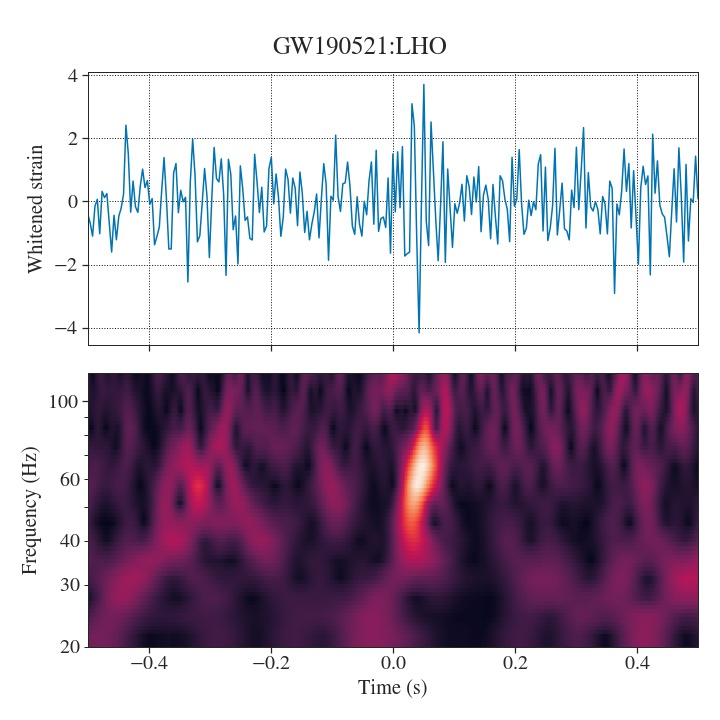}}
      {\includegraphics[width=0.52\linewidth]{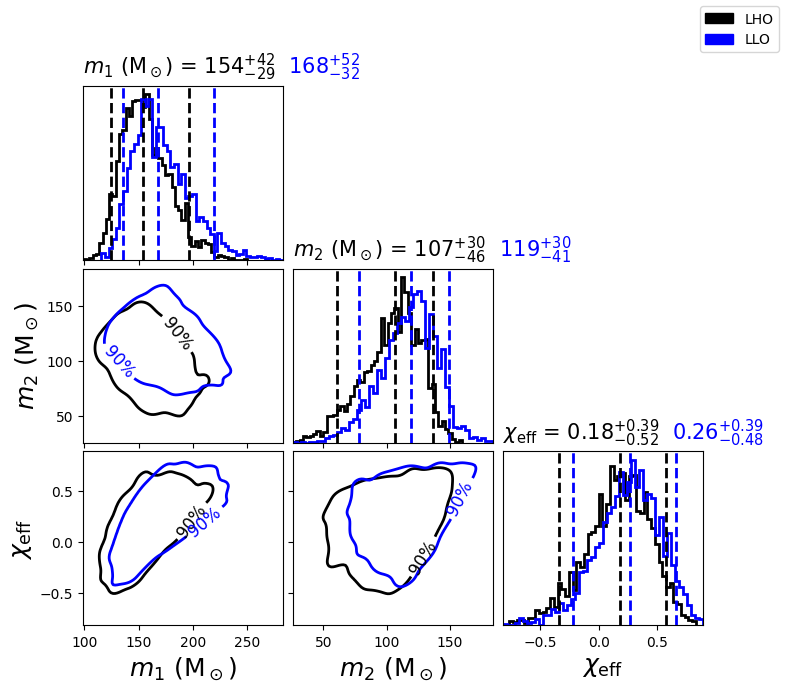}}
      \caption{The top figure in the left panel shows the whitened data in LHO for GW190521. The bottom figure in the left panel shows the Q-transform of the LHO data. The right panel shows the posteriors for LHO and LLO data, which are largely overlapping.}
    \label{fig:gw190521}
    
\end{minipage}
\end{figure*} 

\begin{figure*}
    \begin{minipage}{\textwidth}
      {\includegraphics[width=0.45\linewidth]{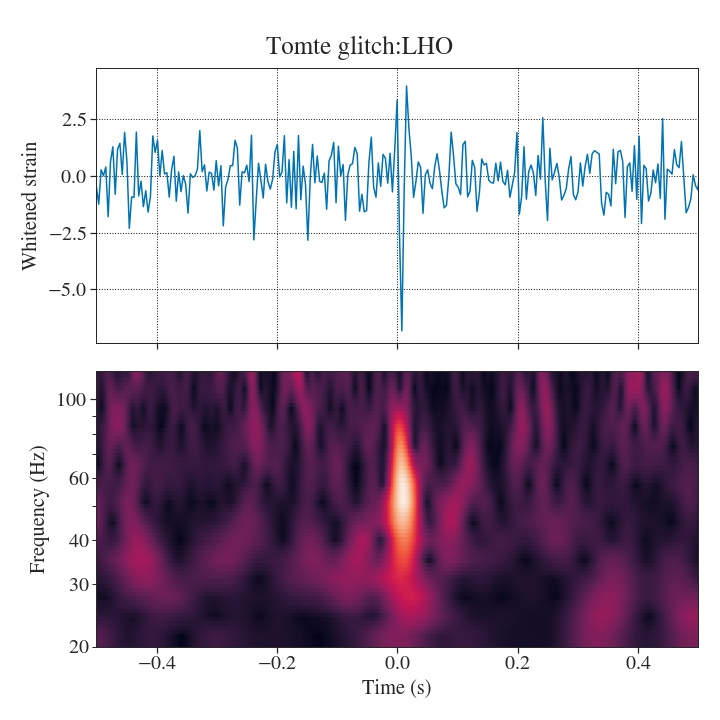}}
      {\includegraphics[width=0.52\linewidth]{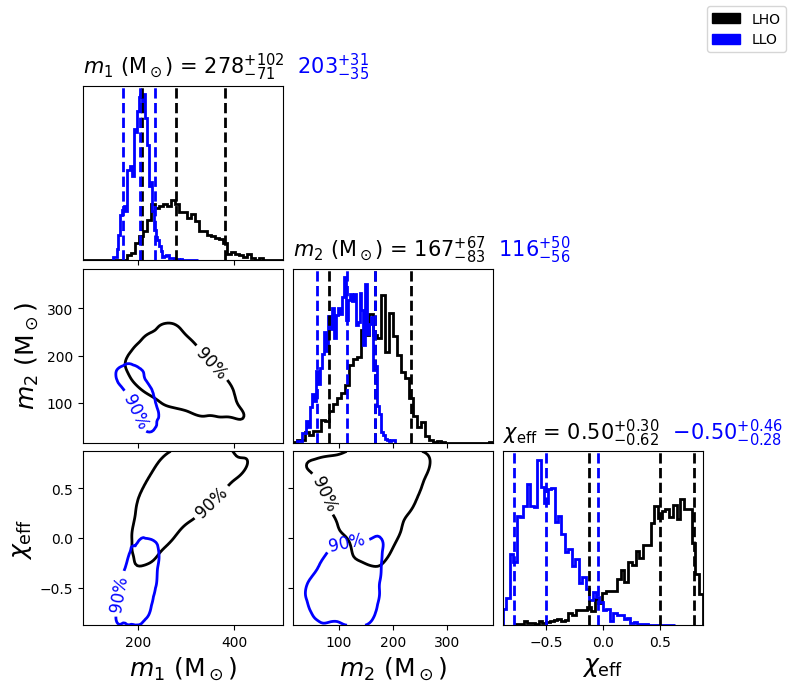}}
      \caption{The top figure in the left panel shows the whitened data around the time of occurrence of a Tomte glitch in LHO. The bottom figure in the left panel shows the Q-transform. The right panel shows the posteriors for LHO and LLO data of the Tomte glitch shown in the left panel and its counterpart in LLO, which triggered the same template. These posteriors are minimally overlapping.}
    \label{fig:tomte}
    
\end{minipage}
\end{figure*} 

As described in section \ref{sec:pe}, the intrinsic parameters determine the phase evolution of the signal. The temporal morphology of the signal depends on the phase evolution. For a quasi-circular waveform model (such as IMRPhenomXAS), the relevant intrinsic parameters are the detector-frame component masses $m_1^\mathrm{det}$ and $m_2^\mathrm{det}$ (which are related to the source-frame masses $m_1^\mathrm{src}$ and $m_2^\mathrm{src}$ by $m_{1,2}^\mathrm{det} = m_{1,2}^\mathrm{src}(1+z)$, $z$ being the redshift) and the effective spin parameter $\chi_\mathrm{eff}$ \cite{santamaria,damour-chieff, racine} which quantifies the degree of alignment of the component spins along the direction of the orbital angular momentum. $\chi_\mathrm{eff}$ is defined as follows:
\begin{equation}
    \chi_\mathrm{eff} = \frac{m_1^\mathrm{det} \chi_{1z} + m_2^\mathrm{det} \chi_{2z}}{m_1^\mathrm{det}+m_2^\mathrm{det}} \, ,
\end{equation}
where $\chi_{1z}$ and $\chi_{2z}$ are the components of the spins along the direction of the orbital angular momentum.

Fig. \ref{fig:gw190521} shows the whitened data, the Q-transform, and the corner plot for the posteriors of $m_1^\mathrm{det}, m_2^\mathrm{det}$ and $\chi_\mathrm{eff}$ using the LHO and LLO data for GW190521. Fig. \ref{fig:tomte} shows the same for a typical Tomte glitch. The duration, frequency range and overall morphologies of the signal and the glitches look similar. However, the posteriors corresponding to LHO and LLO are consistent for the signal (GW190521) and minimally overlap for the glitches.

\subsection{Test Statistic $\mathcal{R}(\mathcal{J})$}{\label{sec:test-statistic}}
In this subsection, we construct a test statistic that ranks candidates based on the consistency of the posteriors across the detector network. For each simulated signal and each pair of background glitches, we calculate $\mathrm{JSD}_{\theta_1,\theta_2}^\mathrm{2D}(p_H||p_L)$, where $p_H$ and $p_L$ are the posteriors for LHO and LLO data respectively. We take all possible combinations of the parameters - $(m_1^\mathrm{det},m_2^\mathrm{det})$, $(m_1^\mathrm{det},\chi_\mathrm{eff})$ and $(m_2^\mathrm{det},\chi_\mathrm{eff})$. Then, we construct their mean $\mathcal{J}$ as: 
\begin{equation}{\label{jsd-mean-defn}}
\begin{split}
    \mathcal{J} = \frac{1}{3} [\mathrm{JSD}^\mathrm{2D}(m_1^\mathrm{det},m_2^\mathrm{det}) + \mathrm{JSD}^\mathrm{2D} (m_1^\mathrm{det},\chi_\mathrm{eff}) + 
    \\ 
    \mathrm{JSD}^\mathrm{2D}(m_2^\mathrm{det},\chi_\mathrm{eff})]
\end{split}
\end{equation}
Due to redundancy, we drop the symbol $p_H||p_L$ and shift the subscript of Eq. \eqref{jsd-2d-defn} (which consists of the parameters for which the JSD is calculated) inside the parentheses. 

\begin{figure*}
    \begin{minipage}{\textwidth}
      {\includegraphics[width=0.32\linewidth]{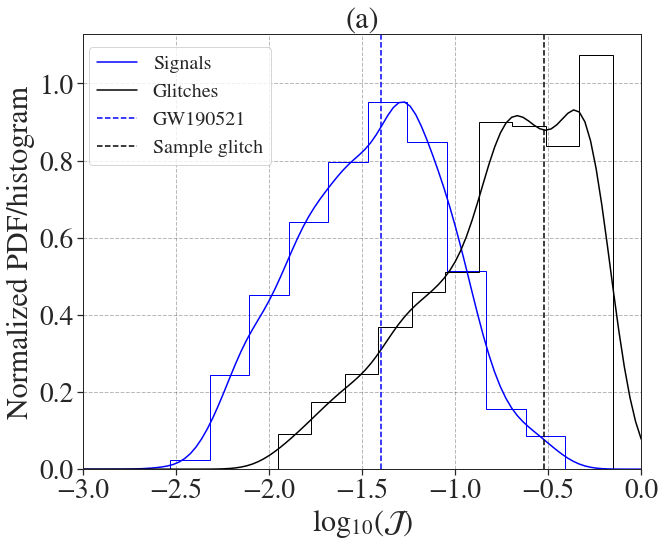}}
      {\includegraphics[width=0.32\linewidth]{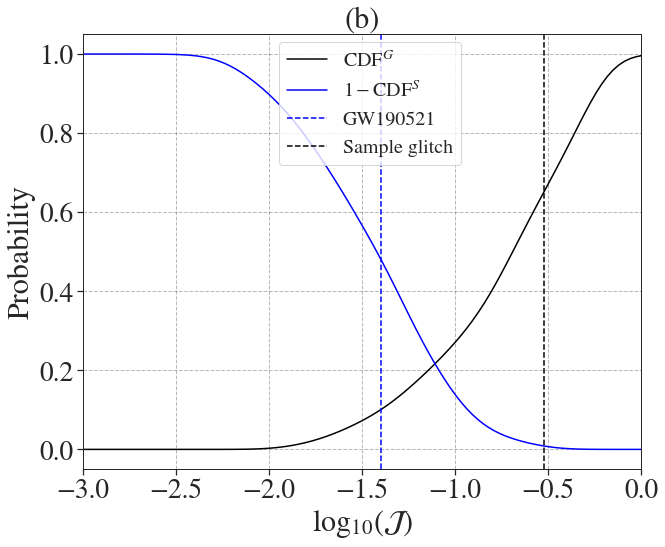}}
      {\includegraphics[width=0.32\linewidth]{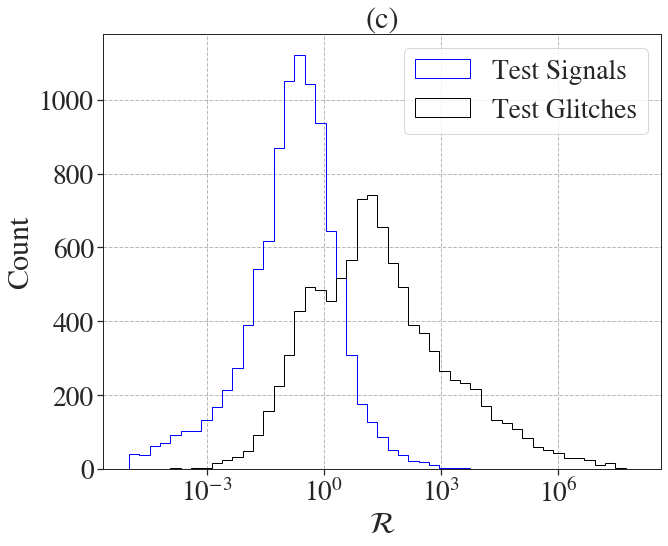}}
      \caption{Panel (a)  shows the normalised histograms of $\mathcal{J}$ for signals (blue stepped) and glitches (black stepped), along with their \acp{KDE} (blue solid for signals and black solid for glitches). Panel (b)  shows 1 - CDF (blue solid) for the signal distribution and CDF (black solid) for the glitch distribution. Panel (c)  shows the distributions of the $\mathcal{R}$ statistic for signals and glitches constructed from the $\mathcal{J}$ distributions shown in panel (a). In panels (a) and (b), the vertical blue dashed line shows the $\mathcal{J}$ value of GW190521 and the vertical black dashed line shows the $\mathcal{J}$ value of the pair of coincident glitches shown in Fig. \ref{fig:tomte}.}
    \label{fig:JSD_2D_mean}
    
\end{minipage}
\end{figure*}

We compute $\mathcal{J}$ as defined in Eq. \eqref{jsd-mean-defn} for all simulated signals and background glitches. Panel (a) of Fig. \ref{fig:JSD_2D_mean} shows the $\mathcal{J}$ distributions for signals and glitches. We observe that a large number of signals have lower $\mathcal{J}$ values than glitches, although there is some overlap between the two distributions.

Here is the scenario in which we are working: We assume that a trigger with $\mathcal{J} = \mathcal{J}_\mathrm{test}$ is already flagged as a GW event with some significance in terms of the \ac{FAR}. The event may not have a very high significance value and may not have been detected in all the searches, like the 200114\_020818 event. In such a scenario, the null hypothesis is that the trigger is of astrophysical origin ($\mathcal{H}_\mathrm{S}$). The hypothesis that the trigger is of instrumental origin ($\mathcal{H}_\mathrm{G}$) is the alternative hypothesis. We note that $1-\mathrm{CDF}^S$ \footnote{$\mathrm{CDF}$ denotes cumulative density function} is the right-tailed p-value for
the signal distribution and will signify the probability of the event being consistent with the null hypothesis. Likewise, $\mathrm{CDF}^G$ is the left-tailed p-value for the glitch distribution. Then, high $1-\mathrm{CDF}^S(\mathcal{J} = \mathcal{J}_\mathrm{test})$ and low $\mathrm{CDF}^G(\mathcal{J} = \mathcal{J}_\mathrm{test})$ imply strong evidence in favour of $\mathcal{H}_\mathrm{S}$ and against ($\mathcal{H}_\mathrm{G}$), and the reverse is true for low $1-\mathrm{CDF}^S(\mathcal{J} = \mathcal{J}_\mathrm{test})$ and high $\mathrm{CDF}^G(\mathcal{J} = \mathcal{J}_\mathrm{test})$. Panel (b)  in Fig. \ref{fig:JSD_2D_mean} shows $1-\mathrm{CDF}^S$ for the signal distribution and $\mathrm{CDF}^G$ for the glitch distribution. Based on the above rationale, we define a test statistic that quantifies the evidence in favour of rejecting $\mathcal{H}_\mathrm{S}$ as follows:   
\begin{equation}{\label{R-defn}}
    \mathcal{R}(\mathcal{J}) = \frac{\mathrm{CDF}^G(\mathcal{J})}{1 - \mathrm{CDF}^S(\mathcal{J})} \\,  
\end{equation}

By construction, $\mathcal{R} \gg 1$ implies that the candidate is more likely to be of instrumental origin, whereas $\mathcal{R} \ll 1$ implies that the candidate is more likely to be of astrophysical origin.

\begin{table*}
    \begin{tabular}{|l|l|l|l|l|l|l|l|l|l|l|} 
    \hline Name & Network SNR & FAR ($\mathrm{yr}^{-1}$) &  $\mathcal{J}$  & $\mathcal{R}$  & p-value\\
    
    \hline  GW150914 &  26.0 &  $\leq$ $10^{-7}$ &   0.009 & 0.002  & 0.921 \\
    
    \hline  GW190519\_153544 &  15.9 & $\leq$ $10^{-5}$ &  0.011  &  0.005  & 0.882 \\

    \hline  GW200224\_222234 &  20.0  & $\leq$ $10^{-5}$ &   0.016  & 0.022 & 0.793 \\

    \hline  GW190727\_060333 &  11.7  & $\leq$ $10^{-5}$ & 0.018   & 0.032 & 0.755 \\

    \hline  GW190503\_185404 &    12.2  & $\leq$ $10^{-5}$ &   0.018  &  0.032 & 0.758 \\

    \hline  GW190521\_074359 &  25.9  & $\leq$ $10^{-5}$ &   0.029  & 0.110 & 0.588 \\

    \hline  GW190602\_175927 &    12.1  & $\leq$ $10^{-5}$ &   0.038  &  0.185 & 0.495 \\
    
    \hline  GW191222\_033537 &    12.5 & $\leq$ $10^{-5}$ &   0.045  & 0.268 & 0.430 \\

    \hline  GW190706\_222641 &  13.4  & $5.0 \times 10^{-5}$ &  0.013 & 0.012 & 0.839 \\

    \hline  GW190521 &  14.3 & $2.0 \times 10^{-4}$ &  0.013  & 0.010  & 0.849 \\ 

    \hline  GW190421\_213856 &   10.7  & $7.7 \times 10^{-4}$ & 0.023 &  0.065 & 0.670\\
    
    \hline  GW200219\_094415 &    10.7  &  $9.9 \times 10^{-4}$  &   0.057 &  0.469 & 0.327\\

    \hline  GW190701\_203306 &   11.2  & $5.7 \times 10^{-3}$ &   0.019  & 0.036 & 0.741 \\
 
    \hline  200114\_020818 &  14.5  & $6.0 \times 10^{-2}$   &   0.241  &  31.873 & 0.017 \\
    
    \hline  200214\_224526 &  13.1  & $1.3 \times 10^{-1}$   &   0.355  &  175.549 & 0.005 \\
    
    \hline
    \end{tabular}
\caption{\label{real-events-2}Shows the results of applying the method developed in this work to real events having source -frame total mass $\geq 65 M_\odot$ and network SNR $\geq 10$. \cite{gwtc-1,gwtc-2,gwtc-3}}
\end{table*}

We fit \acp{KDE} to the distributions of $\mathcal{J}$ for signals and glitches as shown in panel (a)  of Fig. \ref{fig:JSD_2D_mean}. With the assumption that real signals and glitches will follow these \acp{KDE} (also shown in panel (a)  of Fig. \ref{fig:JSD_2D_mean}), we generate points from the signal \ac{KDE} and the glitch \ac{KDE} and treat them like our ``test signals'' and ``test glitches''. Panel (c)  of Fig. \ref{fig:JSD_2D_mean} shows the distributions of $\mathcal{R}$ for test signals and test glitches. Appendix \ref{R-vs-LR} explains the relation between $\mathcal{R}$ and the commonly used \acf{LR} statistic, derived from the \acp{KDE} in panel (a)  of Fig. \ref{fig:JSD_2D_mean}. It shows that there is a one-to-one correspondence between $\mathcal{R}$ and \ac{LR}, and the \ac{ROC} plots of these two statistics, which show that they have comparable distinguishing powers. Here, we use the $\mathcal{R}$ statistic.

\section{Application to candidate events}\label{sec:application}

Table \ref{real-events-2} summarises the results of applying our method to events from \acf{GWOSC} with LHO-LLO. We apply our method to 13 GW-tagged events having source-frame total mass more than $65 M_\odot$ (median values) and 2 marginal candidates observed in the \ac{LHO} and \ac{LLO} detectors with network SNR above 10. Although our primary interest is in the \ac{IMBH} mass range ($10^2\mathrm{M_\odot}$ to $10^5\mathrm{M_\odot}$), we extend the lower mass limit to $ 65 \mathrm{M_\odot}$ to test the robustness of our method. 

We calculate the p-values computed from the distribution of $\mathcal{R}$. Significant GW events are expected to have $\mathcal{R} < 1$, and high p-values. Coincident glitches mimicking signals will have $\mathcal{R} > 1$ and low p-values. 

GW190521, the only significant \ac{IMBH} binary, has an $\mathcal{R}$ value of 0.01 and a p-value of 0.85, which clearly supports $\mathcal{H}_\mathrm{S}$. The marginal candidate 200214\_224526 has an $\mathcal{R}$ value of 176, with a p-value of 0.005. The low p-value is consistent with the instrumental origin of this event. The second marginal candidate 200114\_020818 has $\mathcal{R}$ value of 32 and a p-value of 0.017, which indicates that this candidate is less likely to be of astrophysical origin. \textcolor{blue}{The posteriors of 200114\_020818 and 200214\_224526 are shown in Appendix \ref{marg-cand-post}.}

\textcolor{blue}{Please note that here we assume a 2-detector network with constituent detectors having comparable sensitivities.  However, there may be more than two detectors in the network, and the different detector pairs may not have comparable sensitivities. In that case,  one needs to 
consider different detector pairs and carry out simulations for each pair. Otherwise one may draw incorrect inferences. 
To give an example,  we consider the following LLO-Virgo O3 events  --- GW190620\_030421 (SNR = 10.9, FAR=$0.01 \mathrm{yr}^{-1}$) and GW190910\_112807 (SNR = 13.4, FAR=$0.003 \mathrm{yr}^{-1}$). Most of the network SNR of these events was from LLO detector alone, the Virgo SNR was very low,  practically making them single-detector events.  The p-values from the LHO-LLO distribution are 0.024 and 0.031 indicating them to be not of astrophysical origin.  Clearly, looking at their SNR as well as FAR values,  such a conclusion would be incorrect. This clearly indicates that one should compute the $\mathcal{J}$ and $\mathcal{R}$ distributions for each pair of detectors before applying this method.  A detailed discussion can be found in Appendix \ref{application-to-H1L1V1}.}

\begin{figure*}
    \begin{minipage}{\textwidth}
    {\includegraphics[width=0.32\linewidth]{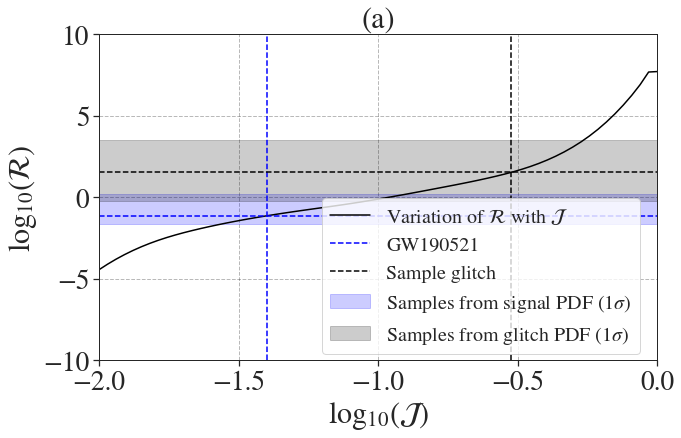}}
    {\includegraphics[width=0.32\linewidth]{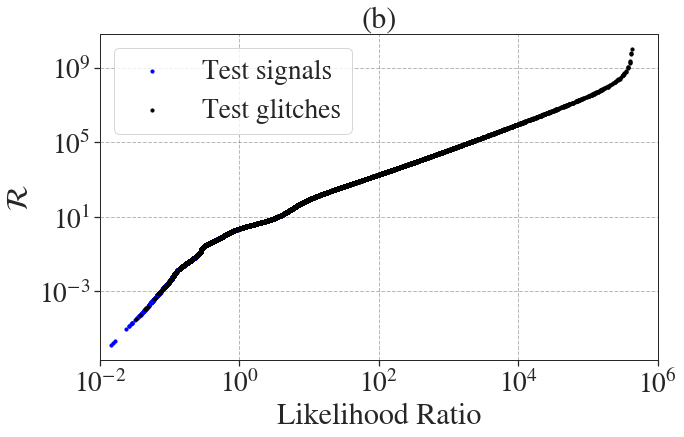}}
    {\includegraphics[width=0.32\linewidth]{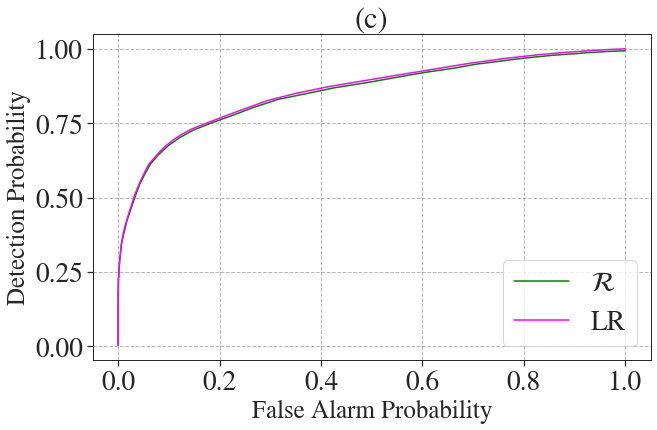}}
    \caption{Panel (a) shows the variation of $\mathcal{R}$ with $\mathcal{J}$. The blue and grey shaded regions represent the $68\%$ confidence intervals of $\mathcal{R}$ for signals and glitches, respectively. Panel (b)  shows the relation between the \ac{LR} and the newly defined $\mathcal{R}$ statistic. $\mathcal{R}$ increases monotonically with \ac{LR}. The blue circles are the test signals, and the black circles are the test glitches. Panel (c)  shows the \ac{ROC} curves of the \ac{LR} and $\mathcal{R}$. The y-axis is the probability of detecting a glitch correctly, and the x-axis is the probability of a signal being misclassified as a glitch.}
    \label{fig:R-vs-LR}   
    \end{minipage}
\end{figure*}

\section{\label{sec:conclusion} Discussion and Conclusion}
Certain classes of noisy glitches mimic the massive
\ac{BBH} events, especially in the \ac{IMBH} binary
parameter space. Several approaches are employed in the
searches such as glitch-specific veto methods, machine learning approaches in the post-processing stage \cite{vetoes,pycbc-paper,chisquared,gravityspy} In this work, we took a new parameter consistency approach to investigate the possibility of loud, short-duration IMBH-like GW triggers being of instrumental origin. Here, we have demonstrated the potential of the 2D \acf{JSD} to distinguish between IMBH binaries and short-duration glitches with comparable SNRs in at least two advanced interferometric detectors. We defined a new test statistic $\mathcal{R}$, synonymous with \acf{LR}, based on 2D JSD to quantify the consistency of posteriors for the intrinsic parameters across the detector network. A pair of coincident triggers giving a low value of $\mathcal{R}$ has a high degree of consistency between the detectors and hence is more likely to be of astrophysical origin. On the other hand, a high value of $\mathcal{R}$ implies a lack of consistency between the detectors, and this indicates that the triggers in question may be of instrumental/environmental origin. We also emphasize that the method
is quite independent of the choice of waveform model. Different waveform models can introduce bias in the estimated parameters. However, if the trigger is of astrophysical origin, the bias in the parameters will be consistent across the detectors, provided the signal SNR is not too
different across the detectors.

We computed the p-values of events to quantify the statistical significance of our classification. We demonstrated this method with the massive \ac{BBH}
merger events and marginal candidates in the observational runs. We observed that all events reported with high significance give low $\mathcal{R}$ values and high p-values as expected. 200214\_224526, whose origin is confirmed to be instrumental, gives a p-value of 0.005.  200114\_020818, which is currently of unknown origin, has a p-value of 0.017, which indicates that this candidate is less likely to be of astrophysical origin. 

Our method can be used in a variety of ways. It can be used in conjunction with detector characterisation to add further confidence in the classification. Further, it can also be folded in the main search pipelines, similar to the $\chi^2$ statistic \cite{chisquared} used in PyCBC-based searches \cite{pycbc-paper,pycbc-imbh} to weigh the candidate events appropriately. Some of these approaches will be taken up in the future.

Similar to JSD, several other ways exist to assess the similarity/dissimilarity between probability distributions. Our work is a proof of concept demonstrating one such metric's use. Other common measures of the dissimilarity between two probability distributions are - Jeffreys distance \cite{jeffreys}, Kolmogorov Smirnov statistic \cite{kolmogorov}, Bhattacharyya distance \cite{bhattacharya} and many more. The exercise that we have performed in this work can be performed with other measures as well which we plan to explore for optimal choice.

\appendix
\section{\label{R-vs-LR} $\mathcal{R}$ and the likelihood ratio}

In this appendix, we show the variation of the  $\mathcal{R}$ statistic with $\mathcal{J}$ and demonstrate that the \ac{LR} statistic and the $\mathcal{R}$ statistic have a one-to-one correspondence and that they have comparable distinguishing powers. 

Panel (a) of Fig. \ref{fig:R-vs-LR} shows the variation of $\mathcal{R}$ with $\mathcal{J}$. As expected, $\mathcal{R}$ increases monotonically with $\mathcal{J}$. In panel (a) of Fig. \ref{fig:JSD_2D_mean}, we plot the \acp{KDE} for signals and glitches.

We generate $10^4$ samples from both the signal \ac{KDE} and the glitch \ac{KDE} and treat them like our ``test signals'' and ``test glitches''. We calculate $\mathcal{R}$ and \ac{LR} for each of these test cases. Panel (b)  of Fig. \ref{fig:R-vs-LR}, shows that the new test statistic $\mathcal{R}$ is a monotonically increasing function of \ac{LR}. A value of \ac{LR} corresponds to a unique value of $\mathcal{R}$ and vice-versa. Panel (c)  of Fig. \ref{fig:R-vs-LR} shows the \acf{ROC} curves for $\mathcal{R}$ and \ac{LR}. We see that they are identical, i.e.,  the distinguishing power of $\mathcal{R}$ is equal to that of \ac{LR}. This indicates that we can use either $\mathcal{R}$ or \ac{LR} as our test statistic. 

\section{\label{marg-cand-post} Posteriors of the marginal candidates 200114\_020818 and 200214\_224526}
\begin{figure*}
    \begin{minipage}{\textwidth}
      {\includegraphics[width=0.45\linewidth]{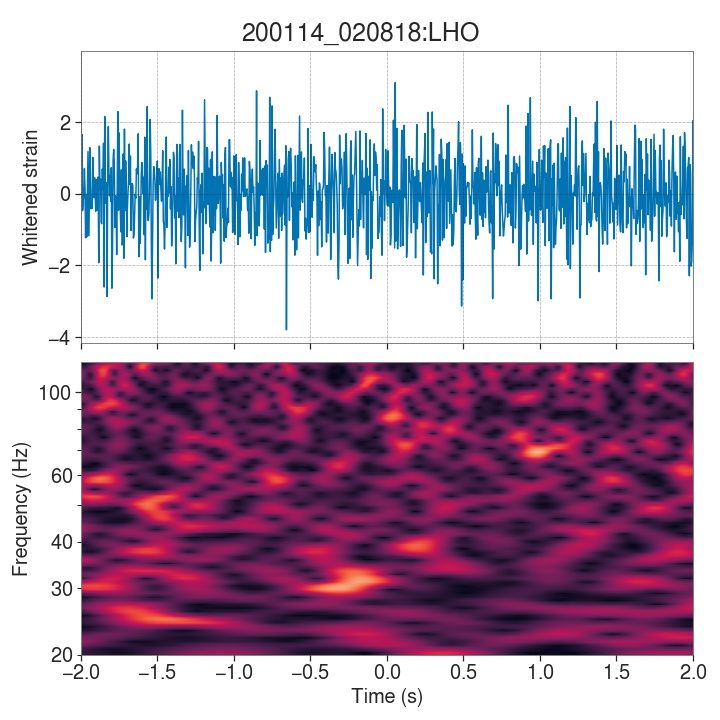}}
      {\includegraphics[width=0.52\linewidth]{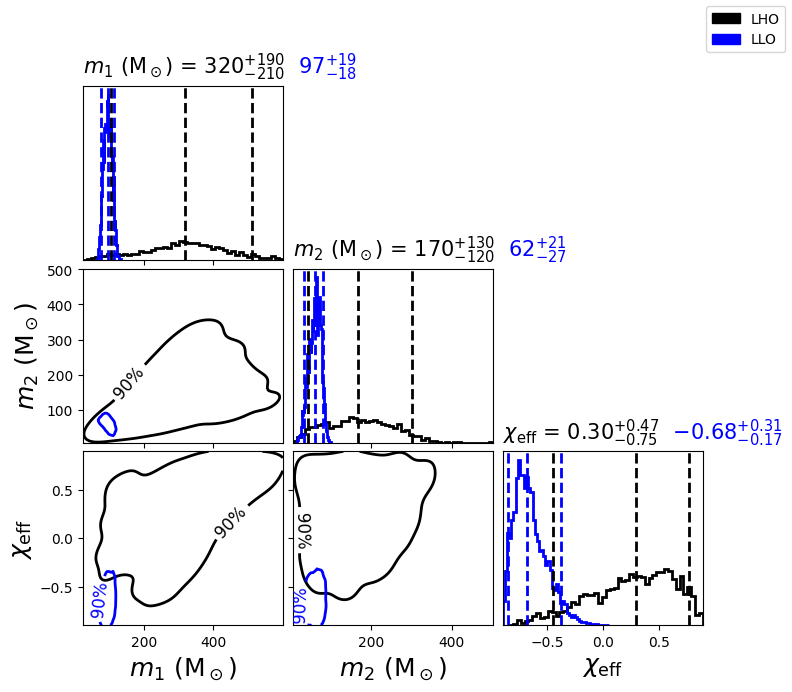}}
      \caption{\textcolor{blue}{The top figure in the left panel shows the whitened data in LHO for 200114\_020818. The bottom figure in the left panel shows the Q-transform of the LHO data. The right panel shows the posteriors for LHO and LLO data.}}
    \label{fig:200114}
    
\end{minipage}
\end{figure*}

\begin{figure*}
    \begin{minipage}{\textwidth}
      {\includegraphics[width=0.45\linewidth]{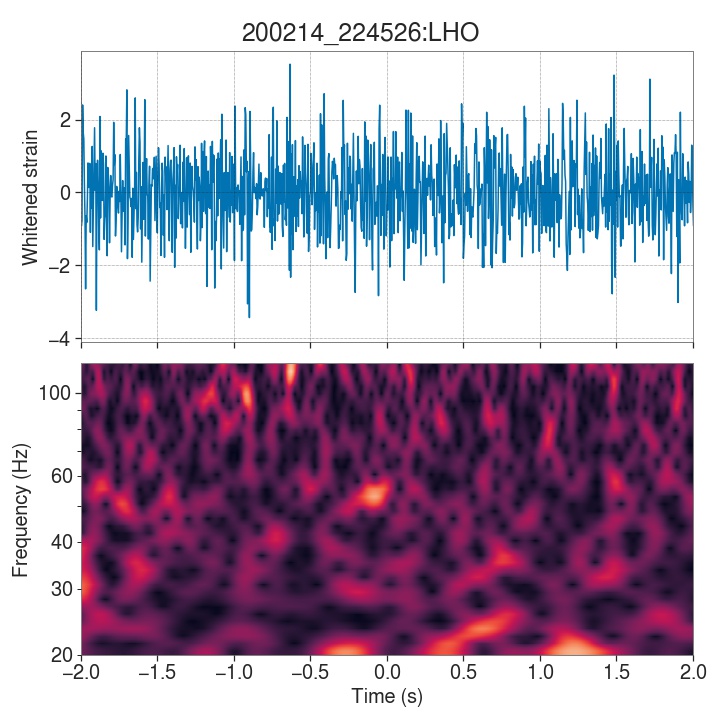}}
      {\includegraphics[width=0.52\linewidth]{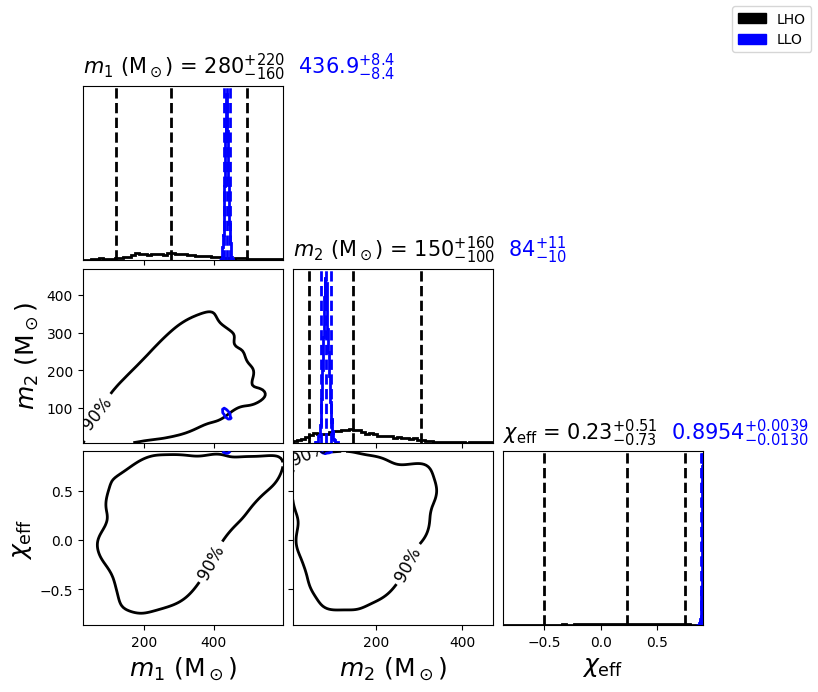}}
      \caption{\textcolor{blue}{The top figure in the left panel shows the whitened data in LHO for 200214\_224526. The bottom figure in the left panel shows the Q-transform of the LHO data. The right panel shows the posteriors for LHO and LLO data.}}
    \label{fig:200214}
    
\end{minipage}
\end{figure*}

\textcolor{blue}
{Figures \ref{fig:200114} and \ref{fig:200214} show the whitened data and the spectrograms of the LHO data for the 2 marginal candidates 200114\_020818 and 200214\_224526, as well as the posteriors in LHO and LLO. For both these candidates, we observe that the posteriors corresponding to the LLO data are sharply localised, indicating that the transients in that detector are well-modelled by \ac{BBH} waveforms described by parameters lying in a small region of the parameter space. Also, for both these candidates, we observe that the posteriors corresponding to the LHO occupy a larger region of the parameter space, indicating that the transients in that detector can be described by a vast range of parameters.}

\section{\label{application-to-H1L1V1} Detectors pairs with different sensitivity}
\textcolor{blue}{
The method introduced in this work was developed for a detector pair with comparable sensitivities.  If more than two detectors are present in the network and have significantly different sensitivities, then the analysis will have to be performed in pairs. For example, if the network has the Virgo detector in addition to the LHO and LLO detectors, then $\mathcal{J}$ and $\mathcal{R}$ will have to be computed for LHO-LLO, LHO-Virgo and LLO-Virgo pairs.} 

\begin{figure*}
    \begin{minipage}{\textwidth}
      {\includegraphics[width=0.45\linewidth]{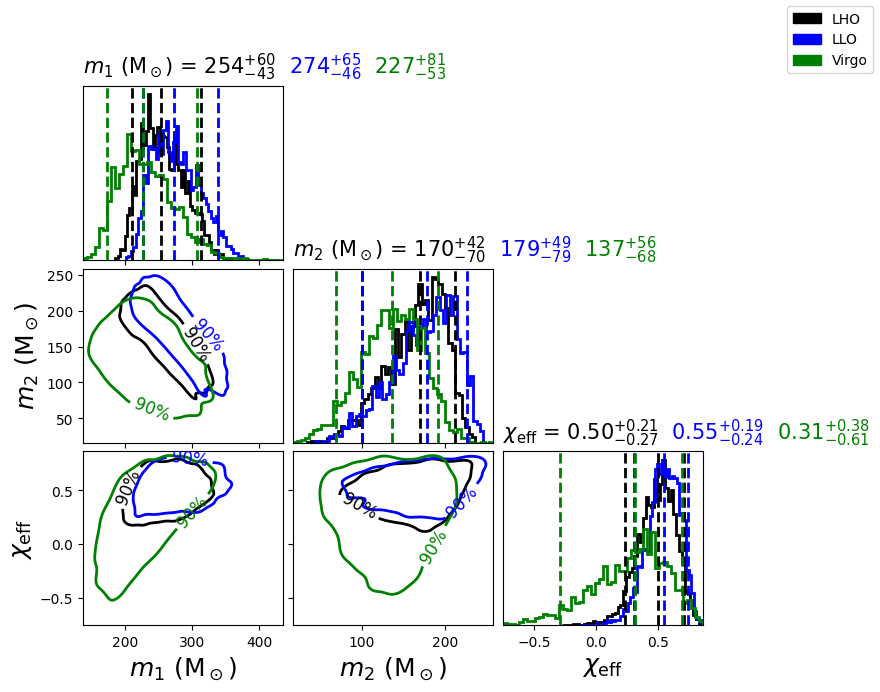}}
      {\includegraphics[width=0.48\linewidth]{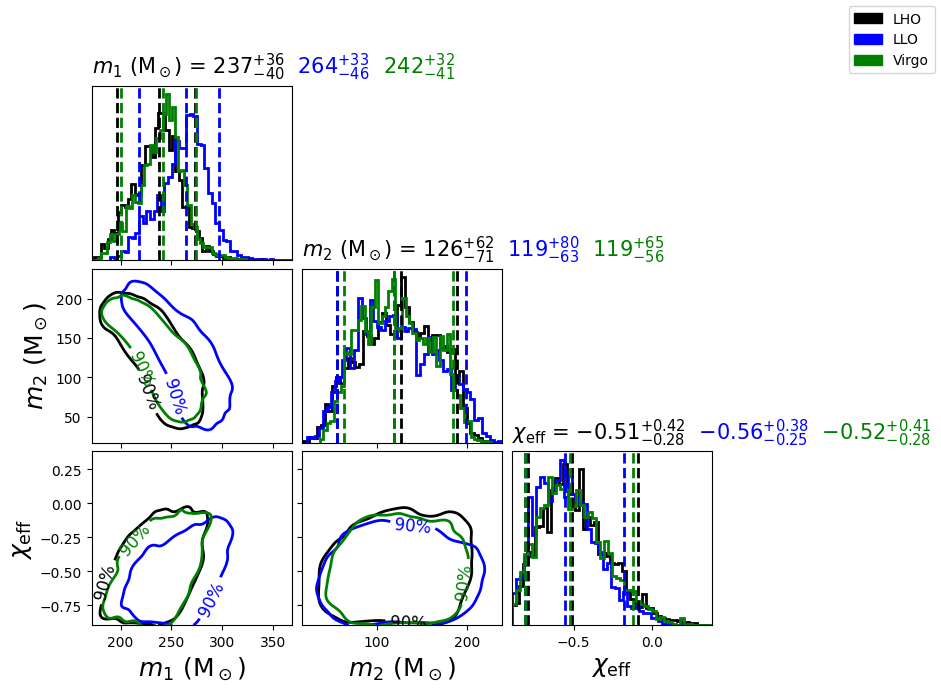}}
      \caption{\textcolor{blue}{The left panel shows the posteriors for a signal injected into LHO, LLO and Virgo with SNRs 16, 18 and 5 respectively. We see that the 2D posteriors in the Virgo detector are much broader than the posteriors in the LHO and LLO detectors. The right panel shows the posteriors for a signal injected into LHO, LLO and Virgo with SNRs 13, 14 and 16 respectively. For this signal, we see that the posteriors have similar widths. }}
    \label{fig:posteriors-H1L1V1}
    
\end{minipage}
\end{figure*}

\textcolor{blue}{
The $\mathcal{J}$ and $\mathcal{R}$ distributions of signals and glitches for a given detector pair may depend upon the relative sensitivities of the two detectors and the nature of the glitches occurring in those detectors. For demonstration, we inject signals into the LHO-LLO-Virgo network, in Gaussian noise coloured by O3 PSDs. We know that the Virgo detector had lower O3 sensitivity than the advanced LIGO detectors, and also has different antenna pattern functions.  Hence, for most sky locations, Virgo SNR will be lower than LHO and LLO SNRs.  As a demonstration,  we take two cases:
one signal injected at a sky location with lower sky sensitivity (due to antenna pattern functions) of Virgo compared to  LHO and LLO, and another signal with sky sensitivity of Virgo comparable to those of LHO and LLO.  This difference in sky sensitivity is reflected in the detector SNRs. The LHO, LLO and Virgo SNRs are 16, 18 and 5 for the first signal and 13, 14 and 16 for the second signal. 
Fig. \ref{fig:posteriors-H1L1V1} shows the posteriors for the two different signals: The left panel shows the first case with high LHO and LLO SNRs but low Virgo SNR, and the right panel shows the injection with comparable SNRs in LHO, LLO and Virgo detectors.  It can be seen that when the SNR difference between the detectors is high, the lower SNR detector has broader 2D posteriors than the high SNR detector. When the SNRs of the detectors are comparable, the 2D posterior widths are also comparable. For the first signal, the $\mathcal{J}$ values for LHO-LLO, LHO-Virgo and LLO-Virgo, are 0.014, 0.041 and 0.06 respectively. Thus, the $\mathcal{J}$ value between Virgo and one of the advanced LIGO detectors is higher than that between the two LIGO detectors, when the Virgo SNR is very low compared to the LHO and LLO SNRs. For the second signal, the $\mathcal{J}$ values for LHO-LLO, LHO-Virgo and LLO-Virgo, are 0.027, 0.011 and 0.034 respectively. In this case also there are variations in the $\mathcal{J}$ values, but this is mainly because the LLO posteriors are slightly different from the LHO and Virgo posteriors. These two figures indicate that if there are large differences in the sensitivities of the two detectors, the $\mathcal{J}$ value tends to be higher, and we thus expect the $\mathcal{J}$ distribution to shift to the right for such detectors. The $\mathcal{R}$ distribution will depend on both the $\mathcal{J}$ distribution for signals and $\mathcal{J}$ distribution for glitches. If the $\mathcal{J}$ distribution for signals moves closer to the $\mathcal{J}$ distribution for glitches, then the $\mathcal{R}$ statistic will have lesser distinguishing power for that pair of detectors.}   

\begin{acknowledgments}
We thank Kipp Cannon, Edoardo Milotti, Florent Robinet, Thomas Dent and Marco Drago for their valuable comments. This research has made use of data, software, and/or web tools obtained from the Gravitational Wave Open Science Center (\href{https://www.gw-openscience.org}{\texttt{https://www.gw-openscience.org}}), a service of LIGO Laboratory, the LIGO Scientific Collaboration and the Virgo Collaboration. Virgo is funded by the French Centre National de Recherche Scientifique (CNRS), the Italian Istituto Nazionale della Fisica Nucleare (INFN), and the Dutch Nikhef, with contributions by Polish and Hungarian institutes. This material is based upon work supported by NSF’s LIGO Laboratory, which is a major facility fully funded by the National Science Foundation. The authors are grateful for computational resources provided by the LIGO Laboratory and supported by National Science Foundation Grants PHY0757058 and PHY-0823459. \textcolor{blue}{SG and KC acknowledge the MHRD, the Government of India, for the fellowship support.} AP acknowledges the support from SERB-Power fellowship grant SPF/2021/000036, DST, India. \textcolor{blue}{This document has LIGO DCC No LIGO-P2300404}.

\end{acknowledgments}


\bibliography{apssamp}
\end{document}